\documentclass[lettersize,journal]{IEEEtran}
\usepackage{amsmath,amsfonts}
\usepackage{algorithmic}
\usepackage{algorithm}
\usepackage{array}
\usepackage{textcomp}
\usepackage{stfloats}
\usepackage{url}
\usepackage{verbatim}
\usepackage{graphicx}
\usepackage{cite}
\usepackage{multirow}
\usepackage{adjustbox}
\usepackage{svg}
\usepackage{subfigure}
\usepackage[colorlinks,allcolors=black,pdftex]{hyperref}

\begin{document}


\title{Learned Adaptive Kernels for High-Fidelity Image Downscaling}


\author{Piyush Narhari Pise and Sanjay Ghosh, \IEEEmembership{Senior Member, IEEE}
\thanks{This work is supported by the Faculty Start-up Research Grant (FSRG), Indian Institute of Technology Kharagpur, India.}
\thanks{Piyush Narhari Pise and Sanjay Ghosh are with Department of Electrical Engineering, Indian Institute of Technology Kharagpur, WB 721302, India (email: \texttt{ sanjay.ghosh@ee.iitkgp.ac.in}).}
}



\markboth{Under Review}%
{Shell \MakeLowercase{\textit{et al.}}: A Sample Article Using IEEEtran.cls for IEEE Journals}



\maketitle

\begin{abstract}
Image downscaling is a fundamental operation in image processing, crucial for adapting high-resolution content to various display and storage constraints. While classic methods often introduce blurring or aliasing, recent learning-based approaches offer improved adaptivity. However, achieving maximal fidelity against ground-truth low-resolution (LR) images, particularly by accounting for channel-specific characteristics, remains an open challenge. This paper introduces ADK-Net (Adaptive Downscaling Kernel Network), a novel deep convolutional neural network framework for high-fidelity supervised image downscaling. ADK-Net explicitly addresses channel interdependencies by learning to predict spatially-varying, adaptive resampling kernels independently for each pixel and uniquely for each color channel (RGB). The architecture employs a hierarchical design featuring a ResNet-based feature extractor and parallel channel-specific kernel generators, themselves composed of ResNet-based trunk and branch sub-modules, enabling fine-grained kernel prediction. Trained end-to-end using an L1 reconstruction loss against ground-truth LR data, ADK-Net effectively learns the target downscaling transformation. Extensive quantitative and qualitative experiments on standard benchmarks, including the RealSR dataset, demonstrate that ADK-Net establishes a new state-of-the-art in supervised image downscaling, yielding significant improvements in PSNR and SSIM metrics compared to existing learning-based and traditional methods.
\end{abstract}


\section{Introduction}
\IEEEPARstart{T}{he} The ubiquitous nature of high-resolution (HR) digital imaging necessitates effective image downscaling techniques. Reducing spatial resolution is essential for various applications, including efficient transmission and storage, adapting content to diverse display capabilities, and stadardizing inputs for downstream computer vision tasks. The primary challenge in image downscaling lies in minimizing resolution while maximally preserving perceptually relevant details and ensuring fidelity to the original content.

Traditional downscaling approaches, such as bicubic \cite{keys1981cubic} or Lanczos interpolation \cite{mitchell1988reconstruction, duchon1979lanczos}, employ fixed low-pass filtering followed by subsampling. While computationally inexpensive, these methods operate identically across all image regions and color channels, often resulting in blurred outputs due to the attenuation of high-frequency information or ringing artifacts near sharp edges \cite{sun2020learned, weber2016rapid}. More recent algorithmic methods attempt content-adaptive filtering \cite{sun2020learned} or optimization-based formulations \cite{liu2017l_} to better preserve structure, but may rely on complex heuristics or struggle with diverse image content and computational efficiency.

Deep learning has catalyzed significant advancements in image restoration tasks, including downscaling. Learning-based methods offer the potential to learn complex and spatially variable downscaling operations directly from data. Current approaches can be broadly categorized. Unsupervised methods and SDFlow \cite{sun2024learning}, typically leverage generative models (GANs or flows) to learn a downsampler that matches the statistical distribution of real-world LR images. While successful in generating realistic textures, their objective function does not directly enforce pixel-wise fidelity to a known ground-truth LR mapping, which can limit performance on metrics like Peak Signal-to-Noise Ratio (PSNR) and Structural Similarity (SSIM) \cite{wang2004image}.

Conversely, supervised or task-aware methods utilize paired HR-LR datasets \cite{cai2019toward} to train downscaling networks optimized for a specific objective, such as minimizing the reconstruction error of a paired super-resolution network \cite{son2021toward, son2021toward} or, more directly, minimizing the error between the generated LR image and its ground-truth counterpart \cite{sun2020learned, niu2022perceptual}. Several methods learn adaptive resampling kernels \cite{xing2023scale, ghosh2023image} or employ invertible architectures \cite{xing2023scale} to model the downscaling process. However, a key limitation persists, many existing kernel-prediction methods derive kernels from shared features or luminance information, neglecting potentially crucial channel-specific degradation characteristics or interdependencies. Furthermore, the network components responsible for kernel prediction are often architecturally simple, potentially limiting their capacity to model highly complex, content-dependent resampling.

In this paper, we address the problem of high-fidelity supervised image downscaling. We introduce ADK-Net (Adaptive Downscaling Kernel Network), a novel deep learning framework specifically designed to maximize the fidelity of the downscaled LR image relative to a ground-truth reference. The central innovation of ADK-Net is the prediction of resampling kernels that are adaptive both spatially (per-pixel) and chromatically (per-channel). By generating distinct kernels for the R, G, and B channels at each pixel location, ADK-Net can model nuanced channel-dependent aspects of the downscaling process, which is hypothesized to be crucial for achieving maximal reconstruction accuracy.

ADK-Net is realized through a carefully designed hierarchical architecture employing ResNet-based components known for their strong representational power \cite{he2016deep}. It consists of a feature extractor preserving spatial detail, a feature downsampling module, and three parallel, channel-dedicated kernel generators. Each generator utilizes sequential ResNet-based Trunk and Branch modules to derive the per-pixel kernels for its specific channel from the shared, downsampled features. The predicted kernels undergo a two-stage normalization (Min-Max scaling followed by sum-to-one normalization) before being applied via weighted resampling to the HR input. The entire network is trained end-to-end by minimizing the L1 distance between the generated LR image and the ground-truth LR image, primarily leveraging the RealSR dataset.

\subsection{Our main contributions are threefold:}
\begin{enumerate}
    \item\noindent We propose ADK-Net, a novel framework for supervised image downscaling based on per-pixel, per-channel adaptive kernel prediction.
    \item\noindent We introduce a specific ResNet-based hierarchical architecture for both feature extraction and kernel generation (Trunk and Branch modules) tailored for this task.
    \item\noindent We demonstrate through extensive experiments that ADK-Net achieves state-of-the-art results on standard benchmarks for supervised image downscaling, significantly outperforming prior methods in terms of PSNR and SSIM against ground-truth LR data.
\end{enumerate}

The remainder of this paper is structured as follows. Section II discusses related work in more detail. Section III presents the architecture and components of the proposed ADK-Net. Section IV details the experimental setup, presents quantitative and qualitative results, and includes ablation studies. Finally, Section V concludes the paper.

\section{Related Work}

Image downscaling, the process of reducing the spatial resolution of an image, has been extensively studied, driven by practical needs in data storage, transmission, and display adaptation. Methodologies range from traditional signal processing techniques to modern deep learning-based approaches. This section reviews pertinent literature, categorizing existing methods to contextualize the contributions of our proposed ADK-Net.

\subsection{Traditional Image Downscaling Methods}
Classical approaches to image downscaling are predominantly rooted in signal processing principles, primarily aiming to mitigate aliasing artifacts introduced during subsampling \cite{shannon1949communication}. These methods typically involve low-pass filtering prior to sampling. Widely adopted linear filters \cite{wolberg1990digital} include Bilinear, Bicubic \cite{smith1981bilinear, mitchell1988reconstruction} and Lanczos \cite{duchon1979lanczos}, which offer computational efficiency but apply a fixed, spatially invariant kernel across the entire image. This content-agnostic nature often leads to a compromise. Aggressive filtering prevents aliasing but causes excessive blurring and loss of fine details, while milder filtering retains some sharpness at the cost of potential aliasing \cite{kopf2013content, weber2016rapid}.

Subsequent efforts focused on designing filters or optimization strategies that better preserve perceptual quality. Content-adaptive filtering techniques, such as the work by Kopf et al. \cite{kopf2013content}, adapt the shape and location of resampling kernels based on local image features, often inspired by bilateral filtering concepts \cite{tomasi1998bilateral, ghosh2019fastscale,ghosh2019saliency,ghosh2019fast}. Optimization-based methods formulate downscaling as a problem of minimizing an objective function, such as structural similarity (SSIM) \cite{wang2004image} or L0-regularized objectives targeting edge preservation \cite{liu2017l_}. For instance, Oztireli and Gross \cite{oeztireli2015perceptually} optimized the downscaled image directly against the original using SSIM, while Liu et al. \cite{liu2017l_} used an L0 prior on gradients to preserve salient edges. Conventional image downscaling algorithms prioritize visual quality, often neglecting the performance impact on downstream computer vision tasks. To address this, Zhang et al. \cite{zhang2011interpolation} introduced a task-guided approach that treats downscaling as the inverse of upsampling, thereby improving the quality of interpolated images derived from the downscaled counterparts. Weber et al. \cite{weber2016rapid} proposed favoring pixels that deviate from their local neighborhood. Other works explored spectral remapping \cite{gastal2017spectral} or co-occurrence statistics \cite{ghosh2023image} to guide the downscaling process. While these methods often yield perceptually superior results compared to linear filtering, they may involve iterative optimization, rely on handcrafted priors, or struggle to generalize across diverse image content and downstream tasks, such as super-resolution. Occorsio et al. \cite{occorsio2022lagrange} introduced a kernel  approximation based method where the core idea was to approximate the downscaled image from the input  by means of global interpolation processes based on (tensor product)  Chebyshev grids of I kind. Further, Vall´ee Poussin type polynomial interpolation based downscaling was proposed in \cite{occorsio2023image}.

\subsection{Learning-Based Image Downscaling}
With the advent of deep learning, research has shifted towards learning downscaling operators directly from data, offering greater flexibility and adaptability. These methods can be broadly classified based on their learning paradigm and objective.

\textit{1) Unsupervised Approaches:}
These methods aim to learn realistic downscaling models without requiring paired HR-LR training data. They often focus on matching the distribution of generated LR images to a target distribution of real-world LR images. 
ADL (Adaptive Downsampling Models) \cite{son2021toward} employs a GAN framework, training a downsampler adversarially against a discriminator to produce realistic LR outputs, complemented by novel loss functions (LFL, ADL) to preserve content and stabilize training. SDFlow \cite{sun2024learning} utilizes normalizing flows to model the conditional distribution of LR images given HR images, allowing for stochastic downscaling by sampling from the learned latent space. While powerful for synthesizing realistic degradations or enabling diverse outputs, these unsupervised methods are not directly optimized for maximizing fidelity against a specific ground-truth LR image, which is the primary goal of our work. Bayesian approaches have also been explored, modeling image priors and degradation statistically \cite{gao2022bayesian, fang2022bayesian}, often requiring complex inference schemes.

\textit{2) Supervised and Task-Aware Approaches:} 
This category encompasses methods trained using paired HR-LR data or optimized jointly with a subsequent task, typically super-resolution (SR). Early works like TAD (Task-Aware Downscaling) \cite{kim2018task} used an autoencoder structure where the encoder performed downscaling and the decoder performed SR, trained jointly to maximize SR performance. Similarly, CR (Compact-Resolution) \cite{li2018learning} trained a downscaling network (CNN-CR) alongside an SR network (CNN-SR), using reconstruction loss and a regularization loss against bicubic downscaling \cite{keys1981cubic} to maintain visual plausibility. BDIS \cite{son2021toward} proposed a balanced dual scaler with "Origin Referenceable Losses" to improve both downscaling and SR quality simultaneously within a joint framework.

More related to our work are methods that explicitly predict resampling kernels. CAR (Content Adaptive Resampler) \cite{sun2020learned} proposed learning non-uniform resampling kernels (weights and offsets) for each LR pixel in an unsupervised manner, guided solely by the reconstruction loss of a paired SR network (EDSR). While achieving good SR restorability, the visual quality of the CAR-downscaled images themselves was secondary. FastDownscaler \cite{niu2022perceptual} presented a lightweight network for efficient downscaling, using simple upsampling losses (bilinear/bicubic) and distillation to balance visual quality and SR restorability. Huang et al. \cite{son2021toward} proposed a scale-arbitrary downscaler for non-learnable upscaling methods, focusing on adapting features based on the scale factor. Park \cite{park2020edge} used edge guidance for adaptive filtering. These methods demonstrate the potential of learned kernels but typically derive them from shared channel features or lack the architectural complexity for modeling highly fine-grained, channel-specific adaptations. Crucially, unlike ADK-Net, they do not generate distinct adaptive kernels for each color channel optimized directly for supervised L1 fidelity against a ground-truth LR image.

\textit{3) Invertible and Scale-Arbitrary Methods:}
A distinct line of research focuses on invertible image rescaling \cite{xiao2020invertible, sun2024learning} or scale-arbitrary downscaling/upscaling \cite{xing2023scale, son2021toward}. IRN \cite{xiao2020invertible} and AIDN \cite{xing2023scale} use invertible neural networks (INNs) and conditional resampling modules, respectively, to handle arbitrary scale factors within a single model and allow mathematically lossless reconstruction (often by hiding information in a latent variable). While powerful for their specific goals, these methods differ significantly from ADK-Net; they often prioritize invertibility or scale flexibility over achieving the absolute highest PSNR/SSIM for a fixed scale factor in a standard supervised setting, and may produce LR images not intended for direct viewing. Convolutional block design for fractional downsampling in \cite{chen2022convolutional} was used in many practical image and video processing applications. A low-complex  invertible image downscaling model by using latent variable
 within their architecture was showned in \cite{zhang2022enhancing}. Guo at al. \cite{guo2023dinn360} introduced another invertible network which  supports 360 degree image downscaling. Recently, a compression-aware image downscaling method was introduced by Li et al. \cite{li2025lightweight}.

\subsection{Positioning of Our Proposed ADK-Net}
The literature reveals a gap in methods specifically designed for high-fidelity supervised image downscaling with a focus on channel-specific adaptation. While unsupervised methods excel at realism and task-aware methods improve SR restorability, they do not directly optimize for matching a ground-truth LR image with maximal pixel-wise accuracy. Existing supervised kernel-prediction methods often overlook channel-specific processing or employ simpler network designs. ADK-Net addresses these limitations by introducing a novel framework that: 
\begin{enumerate}
    \item\noindent Operates under direct supervision using paired HR-LR data (e.g., RealSR) with an L1 loss objective.
    \item\noindent Predicts adaptive resampling kernels independently for each pixel and each color channel, enabling fine-grained chromatic adaptation.
    \item\noindent Utilizes a powerful ResNet-based architecture for both feature extraction and kernel generation to effectively model complex downscaling transformations.
\end{enumerate}
This unique combination allows ADK-Net to establish a new state-of-the-art in supervised image downscaling fidelity.

\section{Proposed Method}

\begin{figure*}[!t]
\centering
\includegraphics[width=7in]{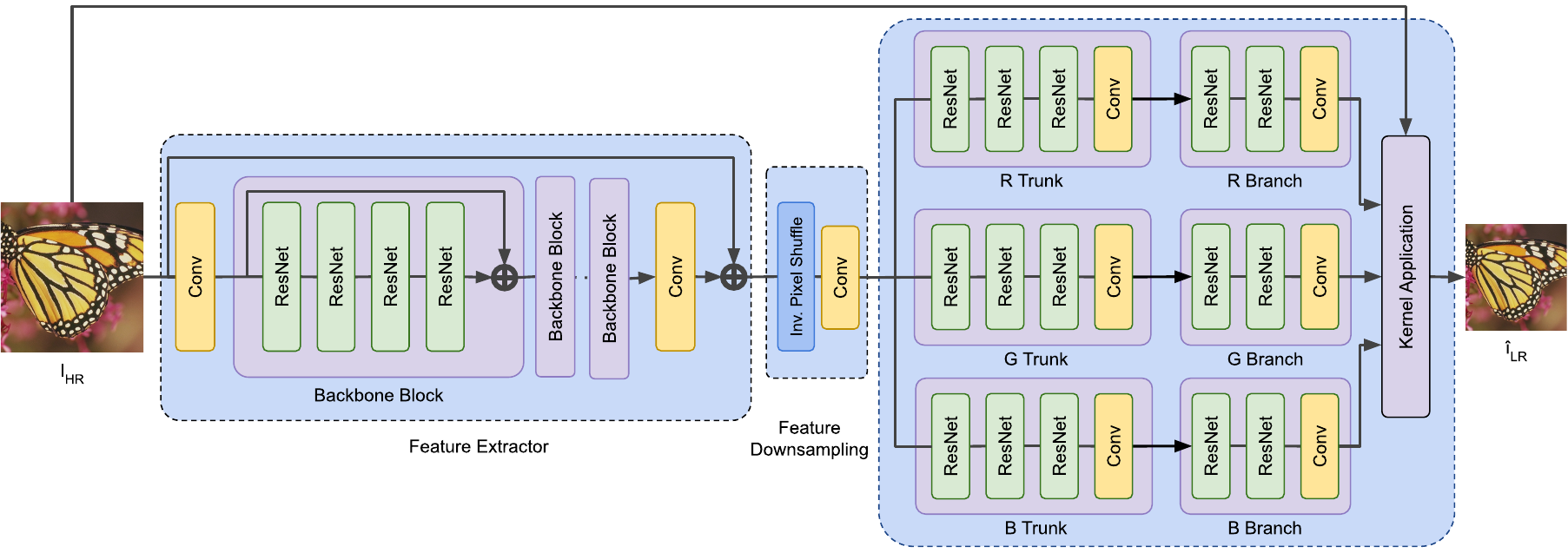}
\caption{Architecture of the proposed Adaptive Downscaling Kernel Network (ADK-Net). The network takes a high-resolution (HR) image \(\mathbf{I_{HR}}\) as input. A ResNet-based Backbone Block first extracts deep features \(F_{HR}\) while preserving spatial resolution. These features are then downsampled to the target low-resolution (LR) spatial grid via an Inverse Pixel Shuffle operation followed by a convolutional layer. The resulting features \(F_{LR}\) feed into three parallel, channel-specific kernel generator streams (for R, G, B channels). Each stream consists of a Channel Trunk (3 ResNet blocks + Conv) to produce channel-specific embeddings \(E_c\) and a subsequent Channel Branch (2 ResNet blocks + Conv) to predict the per-pixel adaptive resampling kernels \(K_{c}\) for that specific channel. Finally, the Kernel Application module applies the normalized predicted kernels to the HR input image to generate the downscaled LR output\({\hat{Y}_{LR}}\).}
\label{fig:architecture}
\end{figure*}

\begin{figure}[!t]
\centering
\includegraphics[width=3in]{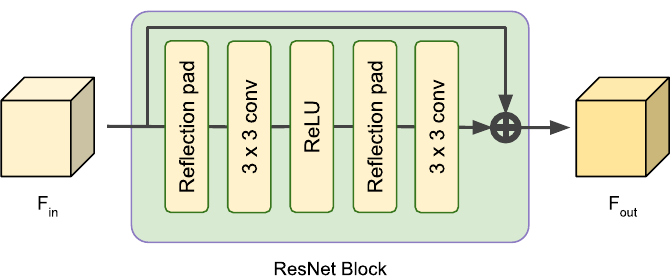}
\caption{The ResNet block architecture utilized in our network. It features two sequential units, each composed of Reflection Padding followed by a 3x3 convolution. A ReLU activation is applied after the first unit. Reflection Padding is employed instead of zero-padding to minimize border artifacts. An identity shortcut connection sums the block's input with the output of the second convolutional layer, enabling the learning of residual functions.}
\label{fig:Resnet}
\end{figure}

In this section, we present the proposed Adaptive Downscaling Kernel Network (ADK-Net), a deep learning framework designed for high-fidelity supervised image downscaling. The core principle of ADK-Net is the prediction of spatially and chromatically adaptive resampling kernels, enabling precise reconstruction of target low-resolution (LR) images from their high-resolution (HR) counterparts.

\subsection{Framework Overview}
The overall architecture of ADK-Net is illustrated in Figure~\ref{fig:architecture}. Given an input HR image  \( \mathbf{I_{HR}} \in \mathbb{R}^{H \times W \times Ch} \) and a target integer scale factor s, ADK-Net generates an LR image \( \mathbf{{I}_{LR}} \in \mathbb{R}^{h \times w \times Ch} \),  where \(h = H/s\) and \(w = W/s\) are the height and width of the output LR image, \(\mathbf{s}\) denotes the downscaling factor and \(Ch\) is the number of channels in the image. The network comprises three main stages: Feature Extraction, Feature Downsampling, and the Per-Channel Kernel Generation and Application. A key characteristic is the parallel processing path within the kernel generator, allowing for independent kernel prediction for each of the R, G, B color channels. The entire network is trained end-to-end using a supervised loss function comparing \(\mathbf{\hat{I}_{LR}}\) to a ground-truth reference \(\mathbf{I_{LR}}\).

\subsection{Feature Extraction}
The purpose of the feature extractor is to derive a rich representation of the input HR image that encapsulates contextual information necessary for predicting appropriate downscaling kernels. It takes \(\mathbf{I_{HR}}\) as input and produces a feature map \( \textbf{F}_{HR} \in \mathbb{R}^{H \times W \times C} \), where \(C \) denotes the number of base feature channels (typically 64). The feature extractor, denoted as \(FE(\cdot)\), consists of:

\textit{Initial Convolution:} A single convolutional layer with Reflection Padding processes the input \(\mathbf{I_{HR}}\) to project it into the initial feature space.

\textit{Backbone Residual Blocks:} A series of residual blocks are employed to progressively refine the features while maintaining spatial resolution. Following the design principles of ResNet \cite{he2016deep}, but adapted for this task, we use a single "Backbone Block". This block itself contains 4 internal ResNet-style residual blocks. Each internal residual block employs a sequence of reflection padding followed by a \(3\times3\) convolution layer followed by ReLU activation function followed by reflection pad followed by a final \(3\times3\) convolution layer as shown in Figure \ref{fig:Resnet} followed by a skip connection adding the block's input to its output. Notably, Batch Normalization layers are omitted to preserve instance-specific details potentially beneficial for adaptive processing \cite{lim2017enhanced}. All convolutions within the backbone maintain \(C\) channels.

\textit{Final Convolution:} A final convolutional layer (\(3\times3\) kernel, \(C\) filters) with Reflection Padding produces the output feature map \(F_{HR} = \phi_{FE}({I_{HR}})\).

Crucially, unlike architectures that progressively downsample features, our extractor preserves the full HR spatial resolution \((H, W)\) throughout this stage.

\subsection{Feature Downsampling}
To align the spatial dimensions of the extracted features with the target LR image grid, a dedicated Feature Downsampling block, is applied to \(F_{HR}\). This block utilizes PixelUnShuffle, the inverse operation of PixelShuffle \cite{shi2016real}, which rearranges elements from an \((H, W, C)\) tensor into an \((H/s, W/s, C\times s^2)\) tensor. This is followed by a \(3\times3\) convolutional layer with Reflection Padding to adjust the channel dimension back to \(C\).
\begin{equation}
F_{LR} = Conv(PixelUnShuffle(F_{HR}))
\end{equation}
The resulting downsampled feature map \(\mathbf{{F}_{LR}} \in \mathbb{R}^{h \times w \times C}\) now contains spatially condensed features corresponding to the LR grid and serves as the input to the kernel generators.

\subsection{Per-Channel Kernel Generation}
This stage is the core of ADK-Net, responsible for predicting the adaptive resampling kernels. Instead of generating a single kernel or deriving kernels from shared features, ADK-Net employs three parallel, independent streams, one for each color channel \(c \in \{R, G, B\} \). Each stream, denoted by \(KG_c(\cdot)\), takes the entire downsampled feature map \(\mathbf{F_{LR}}\) as input and outputs the kernels \(\mathbf{K_{c}} \in \mathbb{R}^{h \times w \times k \times k}\), where \(k\) is the resampling kernel dimensions. Each Kernel Generator stream \(G_k^c\) has an identical architecture but independent weights, consisting of two sequential sub-modules:

\textit{Channel Trunk \((G_T^c(\cdot))\):}
This module transforms the shared features \(\mathbf{F_{LR}}\) into a channel-specific embedding \(\mathbf{E_c} \in \mathbb{R}^{h \times w \times C}\). It is composed of 3 ResNet blocks (identical structure to those in the feature extractor) followed by a final \(3\times3\) convolutional layer.
\begin{equation}
{E_c} = G_T^c({F_{LR}})
\end{equation}

\textit{Channel Branch \((G_B^c(\cdot))\):}
This module takes the channel-specific embedding \(\mathbf{E_c}\) and predicts the final raw kernels for that channel. It consists of 2 ResNet blocks followed by a final \(3\times3\) convolutional layer with \(k \times k\) output channels. This layer effectively maps the \(C\) features to the \(k \times k\) flattened kernel weights for each spatial location.
\begin{equation}
K^{'}_{c} = G_B^c(E_c)    
\end{equation}

Where \(\mathbf{K}^{'}_{c} \in \mathbb{R}^{h\times w\times k\times k} \) represents the generated raw flattened kernels. The use of ResNet \cite{he2016deep} blocks in both Trunk and Branch allows for deep, non-linear transformations, enabling the network to learn complex relationships between image context and the optimal resampling kernel for each channel.

\subsection{Kernel Normalization}
The raw per-channel output kernels \(\mathbf{K}^{'}_c\) require normalization to ensure stable and meaningful resampling. We apply a two-stage normalization process independently to each \(k \times k\) kernel corresponding to every pixel \((x, y)\) in the LR grid and each channel \(c\). Let \(\mathbf{K}^{'}_{c}(x, y)\) be the \(k \times k\) raw kernel reshaped from the flattened output at \((x, y)\) for channel \(c\).

\textit{Min-Max Scaling:}  The kernel values are first scaled to the range \([0, 1]\):
\begin{equation*}
K_{c, scaled}(x,y) = \frac{K^{'}_{c}(x,y) - min(K^{'}_{c}(x,y))}{max(K^{'}_{c}(x,y)) - min(K^{'}_{c}(x,y)) + \epsilon}
\end{equation*}
(where \(\epsilon \) is a small constant for numerical stability).

\textit{Sum-to-One Normalization:} The min-max scaled kernel is then normalized to ensure its elements sum to 1, preserving local energy during resampling:
\begin{equation}
K_c(x, y) = \frac{K_{c, scaled}(x,y)}{sum(K_{c, scaled}(x,y)) + \epsilon}    
\end{equation}
This two-stage process ensures the kernel acts as a valid weighted average operator with non-negative weights.

\subsection{Kernel Application (Downscaling)}
The final LR image \(\mathbf{\hat{I}_{LR}}\) is generated by applying the normalized kernels \(K_c(x, y)\) to the input HR image \(\mathbf{I_{HR}}\). For each output pixel \({\hat{I}_{LR}}(x,\ y,\ c)\) at location \((x,\ y)\) in channel \(c \), we perform a weighted sum over a corresponding patch in the HR image. First, the the location \(x,\ y\) in LR grid is projected to \(u,\ v\) in the HR grid. Where the center coordinates \((u,\ v)\) in the HR grid are calculated:
\[(u,\ v) = (x + 0.5, y+0.5) \times scale - 0.5 \]

Let \(\mathbf{P_{HR}}(u, v, c)\) as the \(k \times k\) patch extracted from \(\mathbf{I_{HR}}\) centered at \((u, v)\) for channel \(c\). The output pixel value is computed by convolving the normalized kernel \(K_c(x,\ y)\) with the HR image patch centered at \((u,\ v)\):

\begin{equation*}
\label{eq:adaptive_kernel_conv}
\hat{I}_{LR}(x,y,c) =  \sum_{i=-\frac{k}{2}}^{\frac{k}{2}}\sum_{j=-\frac{k}{2}}^{\frac{k}{2}} K_c(x, y)(i,\ j) \cdot I_{HR}(u + i, v + j,\ c)
\end{equation*}

\begin{equation}
\hat{I}_{\text{LR}}(x,\ y,\ c) = \sum K_c(x,\ y) \odot P_{\text{HR}}(u,\ v,\ c)
\end{equation}

where indices \(i,\ j\) for the kernel \(K_c(x,y)\) are relative to its center, and \(\mathbf{I_{HR}}(u', v', c)\) refers to the pixel value at (u', v') in channel c of the input HR image. Appropriate boundary handling (Reflection Padding, as used throughout the network) is applied when accessing \(\mathbf{I_{HR}}\) pixels near the image borders. This operation effectively performs adaptive local resampling guided by the learned kernels.

\subsection{Training Objective}
ADK-Net is trained in a supervised manner using paired HR-LR images \((\mathbf{I_{HR}},\ \mathbf{I_{LR}})\). The network parameters \(\theta \) (including weights and biases of all convolutional layers and residual blocks) are optimized by minimizing the L1 loss between the generated LR image \(\mathbf{\hat{I}_{LR}} = {f}(\mathbf{I_{HR}};\ \theta)\) and the ground-truth LR image \(\mathbf{I_{LR}}\):
\begin{equation}
\mathcal{L}(\theta) = \frac{1}{N}\sum_{\mathbf{\hat{p}} \in \mathbf{\mathbf{\hat{Y}_{LR}}}}{||p - \hat{p}||}_1
\label{eq:loss}
\end{equation}

where \(p \in \mathbf{I_{LR}}\) and \(\hat{p} \in \mathbf{\hat{I}_{LR}}\) represent the ground-truth and reconstructed pixel value, \(N\) indicates the number of pixels times the number of color channels. The L1 loss is chosen as it is known to encourage sharpness and is less sensitive to outliers compared to L2 loss, often leading to better perceptual results in image restoration tasks \cite{lim2017enhanced}. The loss is computed over the RGB channels and averaged across all pixels and batch samples.


\section{EXPERIMENTS}

This section details the experimental evaluation of the proposed ADK-Net framework. We first describe the experimental setup, including datasets, evaluation metrics, implementation specifics, and the methods used for comparison. Subsequently, we present quantitative and qualitative results comparing ADK-Net against state-of-the-art methods. Finally, we conduct ablation studies to validate the contributions of key components within our proposed architecture.

\begin{figure*}
    \centering
    \subfigure[Ground-truth LR.]{\includegraphics[width=0.19\textwidth]{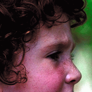}} 
    \subfigure[Predicted LR.]{\includegraphics[width=0.19\textwidth]{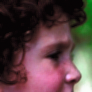}} 
   \subfigure[Red channel kernels.]{\includegraphics[width=0.19\textwidth]{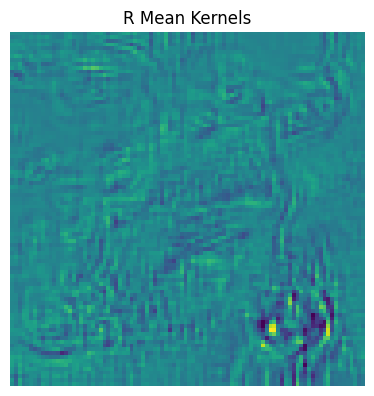}} 
   \subfigure[Green channel kernels.]{\includegraphics[width=0.19\textwidth]{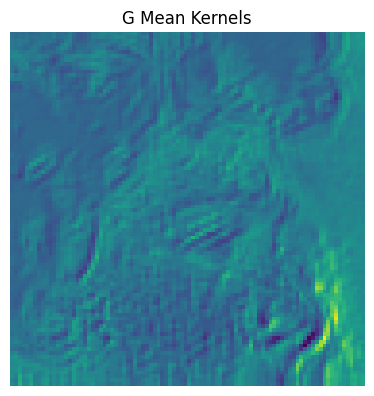}}
    \subfigure[Blue channel kernels.]{\includegraphics[width=0.19\textwidth]{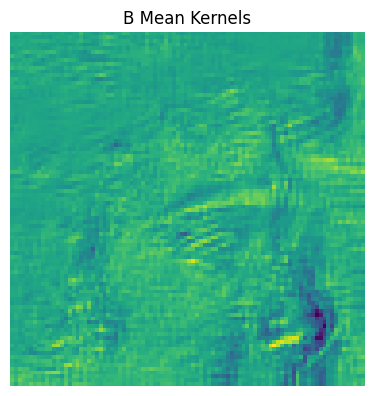}}
\caption{Visual comparison for $3\times$ downscaling. (a) Ground-truth LR image. (b) LR image downscaled by Per-Channel ADK-Net. (c-e) Spatially averaged predicted kernels for Red, Green, and Blue channels, respectively, demonstrating learned adaptability.  Our method attains PSNR/SSIM values 32.79 / 0.916.}
    \label{fig:RGB_kernels}
\end{figure*}
\begin{figure*}
    \centering
    \subfigure[Ground-truth (LR).]{\includegraphics[width=0.24\textwidth]{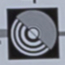}} 
    \subfigure[Bicubic \cite{keys1981cubic}, ($22.93, 0.821$).]{\includegraphics[width=0.24\textwidth]{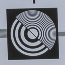}} 
   \subfigure[DPID, ($27.09, 0.885$).]{\includegraphics[width=0.24\textwidth]{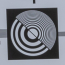}} \\
   \subfigure[IDCL, ($26.19, 0.873$).]{\includegraphics[width=0.24\textwidth]{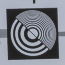}}
    \subfigure[SDFlow, ($33.01, 0.955$).]{\includegraphics[width=0.24\textwidth]{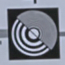}}
    \subfigure[Ours (ADK-Net) \ ($34.11, 0.971$).]{\includegraphics[width=0.24\textwidth]{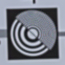}}
\caption{Visual comparison of downscaled HR images via different downscaling methods. (a) Ground Truth LR, (b) Bicubic, (c) DPID, (d) IDCL, (e) SDFlow and (f) Our ADK-Net with scale factors 4x.}
    \label{fig:grey}
\end{figure*}

\subsection{Experimental Setup}
 \textit{1) Datasets:} We utilized the RealSR dataset \cite{cai2019toward} for training our supervised ADK-Net model. RealSR provides high-quality paired HR and corresponding ground-truth LR images captured using a DSLR camera setup, making it suitable for training and evaluating supervised downscaling methods. For testing, we evaluated performance on the RealSR validation set and four standard benchmark datasets commonly used in super-resolution and related tasks: Set5 \cite{bevilacqua2012low}, Set14 \cite{zeyde2010single}, BSD100 \cite{martin2001database}, and Urban100 \cite{huang2015single}. We conducted experiments for integer scale factors \(s = \{2, 3, 4\}\).

\textit{2) Evaluation Metrics:} We quantitatively evaluated the performance using two standard image fidelity metrics: peak signal-to-noise ratio (PSNR) and the structural similarity index measure (SSIM) \cite{wang2004image}. Following common practice in related literature \cite{lim2017enhanced, sun2020learned}, metrics were calculated on both the full RGB image and the Y-channel (luminance) after converting the images to the YCbCr color space. Higher values for both PSNR and SSIM indicate better fidelity of the generated LR image compared to the ground-truth LR image.

\textit{3) Implementation Details:} ADK-Net was implemented using the PyTorch framework \cite{paszke2019pytorch}. Separate models were trained for each scale factor \(s \in \{2, 3, 4\}\). The network architecture employs \(C\) = 64 feature channels. The feature extractor contains one backbone block with 4 internal ResBlocks. The kernel generators use 3 and 2 ResNet blocks in the trunk and branch, respectively. The kernel size was set to \((k,\ k)\), where \(k = 2\times scale + 1\). All convolutional layers used Reflection Padding, and no Batch Normalization layers were used.
Models were trained using the Adam optimizer \cite{kingma2014adam} with an initial learning rate of \(1 \times 10^{-4}\). A dynamic learning rate schedule was employed, reducing the learning rate by half when the validation loss (on a subset of RealSR) plateaued for a certain number of epochs. Training was performed for approximately 100 epochs. Due to memory constraints, training utilized randomly cropped HR patches of size \(192\times192\), \(256\times256\), or \(512\times512\) pixels from the RealSR training set, with corresponding ground-truth LR patches. Data augmentation included random horizontal flips and \(90^\circ,\ 180^\circ,\ 270^\circ\ \) rotations. The batch size typically ranged from 4 to 8 depending on patch size and GPU memory. Training was conducted on an NVIDIA T4 GPU available via Google Colab, with each model taking approximately 1 hour per 100 epochs to train. During inference, the model processes HR images of arbitrary size.

\begin{figure*}
    \centering
    \subfigure[Input (HR).]{\includegraphics[width=0.27\textwidth]{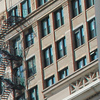}} 
    \subfigure[Bicubic \cite{keys1981cubic}, ($24.91, 0.839$).]{\includegraphics[width=0.27\textwidth]{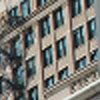}} 
   \subfigure[DPID, ($25.03, 0.832$).]{\includegraphics[width=0.27\textwidth]{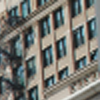}} \\
   \subfigure[IDCL, ($24.84, 0.827$).]{\includegraphics[width=0.27\textwidth]{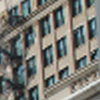}}
    \subfigure[CAR, ($25.00, 0.8297$).]{\includegraphics[width=0.27\textwidth]{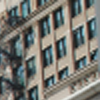}}
    \subfigure[Ours (ADK-Net) \ ($25.54, 0.840$).]{\includegraphics[width=0.27\textwidth]{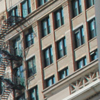}}
\caption{Visual comparisons of the reconstructed HR images produced along with PSNR/SSIM scores by the (c) Bicubic interpolation (BI+BI), (d) IDCL+BI, (e) CAR and (f) Our ADK-Net with scale factors 2x.}
    \label{fig:building}
\end{figure*}

\subsection{Qualitative Analysis With Visual Results}

We first perform an in-depth visual analysis of the kernel-learning capability of our method. In Figure \ref{fig:RGB_kernels}, we show the spatially averaged predicted kernels for the red, green, and blue channels for the output image generated by our method. For reference, the ground-truth image is also shown in Figure \ref{fig:RGB_kernels}. Notice that both red and green channels are able to retain the structures presented in the input.

In Figure \ref{fig:grey}, we provide qualitative comparisons to visually assess the performance of ADK-Net. for scale factor $4\times$.  We display the downscaled results  generated by Bicubic, DPID, IDCL, SDFlow, and the proposed ADK-Net for visual comparion of a grey-scale (challenging) image patch selected from the Urban100 dataset. The original ground-truth (low-resolution i.e. same dimension as the downscaled ones) is also shown Figure \ref{fig:grey} for reference. Notice the high-frequency artifacts in outputs of bicubic, DPID, and IDCL methods. On visual perception, SDFlow result is competitive to ADK-Net. However, our method ADK-Net gives both higher PSNR and SSIM metrics than SDFlow on this image.

\begin{figure*}
    \centering
    \subfigure[Input (HR).]{\includegraphics[width=0.24\textwidth]{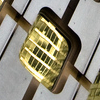}} 
    \subfigure[Bicubic \cite{keys1981cubic}, ($25.01, 0.893$).]{\includegraphics[width=0.24\textwidth]{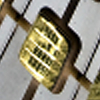}} 
   \subfigure[DPID, ($24.68, 0.887$).]{\includegraphics[width=0.24\textwidth]{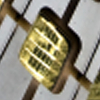}} \\
   \subfigure[IDCL, ($24.63, 0.887$).]{\includegraphics[width=0.24\textwidth]{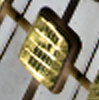}}
    \subfigure[CAR, ($24.56, 0.884$).]{\includegraphics[width=0.24\textwidth]{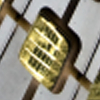}}
    \subfigure[Ours (ADK-Net) \ ($35.69, 0.991$).]{\includegraphics[width=0.24\textwidth]{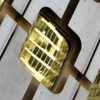}}
\caption{Visual comparisons of the reconstructed HR images produced along with PSNR/SSIM scores by the (c) Bicubic interpolation (BI+BI), (d) IDCL+BI, (e) CAR and (f) Our ADN with scale factors 2x.}
    \label{fig:high-freq}
\end{figure*}

We display visual results on two natural images in Figures \ref{fig:building} and \ref{fig:high-freq} and compare with a state-of-the-art method CAR \cite{}. In this cases, we do not have access to the ground-truth LR images. Therefore, to demonstrated the effectiveness of a downscaling method, we first scale down  the input (high-resolution) image by the respective method and then reconstruct (i.e. super-resolution) using bicubic interpolation. Finally, we use this interpolated (/reconstructed) to visually compare with the input and also to compute PSNR/SSIM. In Figure \ref{fig:building}, we see that the outputs of all existing methods incliding CAR suffer are blurred by the downscaling process. In contrast, our method ADK-Net produces high-quality blur-free output image. Similarly, the blurring effect is also evident in the visual comparion in Figure \ref{fig:high-freq}. It is important to note that the PSNR/SSIM values of the reconstructed images using our method are significantly higher than all other methods. 

Visual inspection confirms the quantitative findings. Bicubic interpolation produces noticeably blurry results, failing to preserve fine textures and sharp edges. Results from SDFlow, while potentially capturing realistic textures, may exhibit artifacts or deviations from the ground-truth structure due to their unsupervised nature and focus on distribution matching rather than pixel-wise fidelity. CAR, while better than bicubic, can sometimes struggle with complex patterns or introduce minor artifacts. In contrast, ADK-Net consistently generates LR images that are significantly sharper and richer in detail compared to all baselines. Fine structures, intricate textures, and clean edges are well-preserved, closely resembling the ground-truth LR appearance. Artifacts commonly associated with downscaling, such as aliasing or ringing, are effectively suppressed. These visual improvements are particularly evident in challenging regions with repetitive patterns or fine lines, demonstrating the effectiveness of the learned per-pixel, per-channel adaptive kernels.




\subsection{Quantitative Analysis Using Metrics}
We present the quantitative performance comparison for scale factors \(s = 2, 3,\) and \(4\) in Tables \ref{tab:table1}, \ref{tab:table2}, and \ref{table:sr_performance_on_div2k} datasets. The tables report PSNR and SSIM values 
averaged over each test dataset. 
In Table \ref{tab:table1}, we train our model on RealSR dataset and also test on same via cross-validation. To study the robustness of our model, we also perform another experiment reported in Table \ref{tab:table2} where the model is trained on RealSR and tested on datasets: Set5, Set14, BSD100, and Urban100. In all cases, we have ground-truth images which are used to PSNR/SSIM computation. Notice that our method ADK-Net outperforms the state-of-the-art method SDFlow \cite{sun2024learning} by big margins. In Table \ref{table:sr_performance_on_div2k}, we compare the bicubic interpolation (BI) performance from different downscaling methods in terms of PSNR/SSIM on DIV2K dataset.

The results clearly demonstrate the effectiveness of the proposed ADK-Net. Across all tested scale factors and datasets, ADK-Net consistently achieves the highest PSNR and SSIM scores, often by a significant margin over the compared methods. For instance, on the RealSR validation set at scale x4, ADK-Net outperforms the second-best method SDFlow \cite{sun2024learning} by significant PSNR and SSIM values. Similar substantial gains are observed on the standard benchmarks like Set14 and Urban100.

\begin{table*}
\label{table:sr_performance_on_div2k}
\caption{Comparison of the Downscaling performance from different downscaling methods in terms of PSNR (dB)/SSIM on RealSR(TEST SET). Best performers are highlighted by bold. Note that DASR \cite{liang2022efficient} acheives PSNR/SSIM as $(32.26 / 0.9472)$ for  $4\times$ scale.}
\centering
\renewcommand{\arraystretch}{1.2}
\begin{tabular}{|c|c|c|c|c|c|}
\hline
\textbf{scale} & \textbf{Bicubic} \cite{keys1981cubic} & \textbf{DPID} [TOG 2016] \cite{weber2016rapid} & \textbf{IDCL} [JVIS 2023]  \cite{ghosh2023image} & \textbf{SDFlow} [TPAMI 2024] \cite{sun2024learning} & \textbf{ADK-Net (OUR)}\\
\hline
2x & 30.28 / 0.9007 & 31.60 / 0.9216 & 31.57 / 0.9209   & 32.82 / 0.938 & \textbf{35.32 / 0.9645}\\
3x & 27.36 / 0.8480 & 29.27 / 0.8909 & 28.77 / 0.8819   & 31.73 / 0.921  & \textbf{33.56 / 0.9550}\\
4x & 25.79 / 0.8168 & 30.30 / 0.8698 & 28.10 / 0.8759  
& 32.55 / 0.9466 & \textbf{33.27 / 0.9560}\\
\hline
\end{tabular}
 \label{tab:table1}
\end{table*}

\begin{table*}
\label{table:sr_performance_on_diff_datasets}
\centering
\caption{Quantitative evaluation results (PSNR / SSIM) of different image downscaling methods for downscaling on benchmark datasets: SET5,
 SET14, BSD100 and URBAN100. For $4\times$, recent method DASR  \cite{liang2022efficient} acheives PSNR/SSIM values: ($29.62 / 0.9326$) for Set5, ($29.72 / 0.9253$) for Set14, ($30.91 / 0.9224$) for BSD100, and ($28.33 / 0.9091$) for URBAN100 datasets.}
\renewcommand{\arraystretch}{1.3}
\begin{tabular}{|c|c|c|c|c|c|c|}
\hline
\textbf{Downscaling} & \multicolumn{2}{c|}{\textbf{Bicubic} \cite{keys1981cubic}}  & \textbf{DPID} [TOG 2016] \cite{weber2016rapid}  & \textbf{IDCL} [JVIS 2023]  \cite{ghosh2023image} & \textbf{SDFlow} [TPAMI 2024] \cite{sun2024learning} & \textbf{ADK-Net (OUR)} \\
\hline
\multirow{3}{*}{Set5} & 2x & 36.20 / 0.9798 & 43.66 / 0.9956 & 40.93 / 0.9913  & 36.62 / 0.981 & \textbf{44.29 / 0.9961}\\
                       & 3x & 30.34 / 0.9417 & 42.20 / 0.9959 & 38.87 / 0.9910 & 31.04 / 0.946 & \textbf{43.92 / 0.9973}\\
                       & 4x & 28.64 / 0.9350 & 41.04 / 0.9956 & 37.65 / 0.9901  & 26.90 / 0.8859 & \textbf{42.75 / 0.9971}\\
\hline
\multirow{3}{*}{Set14}  & 2x & 33.20 / 0.9616 & 40.35 / 0.9908 & 37.04 / 0.9836 & 33.85 / 0.976  & \textbf{42.46 / 0.9939}\\
                        & 3x & 27.21 / 0.8816 & 39.58 / 0.9909 & 35.46 / 0.9810 & 27.86 / 89.952  & \textbf{42.20 / 0.9947}\\
                       & 4x & 26.14 / 0.8757 & 38.92 / 0.9909 & 34.81 / 0.9811 & 26.78 / 0.8516 & \textbf{41.47 / 0.9945}\\ 
\hline
\multirow{3}{*}{BSD100} & 2x & 32.73 / 0.9536 & 40.01 / 0.9889 & 37.74 / 0.9825 & 38.12 / 0.987  & \textbf{42.05 / 0.9927}\\
& 3x & 27.15 / 0.8607 & 39.81 / 0.9889 & 36.68 / 0.9801 & 28.34 / 88.304 & \textbf{42.46 / 0.9935}\\
                      & 4x & 26.34 / 0.8470 & 39.57 / 0.9890 & 36.27 / 0.9799 & 29.81 / 0.9155 & \textbf{42.00 / 0.9932}\\
\hline
\multirow{3}{*}{Urban100} & 2x & 30.22 / 0.9605 & 37.90 / 0.9909 & 34.86 / 0.9839 & 34.90 / 0.863 & \textbf{40.06 / 0.9940}\\
                          & 4x & 23.46 / 0.8470 & 36.20 / 0.9861 & 32.12 / 0.9705 & 27.16 / 0.8988 & \textbf{38.94 / 0.9915}\\
                          
\hline
\end{tabular}
\label{tab:table2}
\end{table*}

\begin{table*}
\caption{Comparison of the Bicubic Upscaling performance from different downscaling methods in terms of PSNR/SSIM on DIV2K dataset. Best performers are highlighted by bold.}
\centering
\renewcommand{\arraystretch}{1.2}
\begin{tabular}{|c|c|c|c|c|c|}
\hline
\textbf{scale} & \textbf{Bicubic} \cite{keys1981cubic} & \textbf{DPID} [TOG 2016] \cite{weber2016rapid}  & \textbf{IDCL} [JVIS 2023]  \cite{ghosh2023image} & \textbf{SDFlow} [TPAMI 2024] \cite{sun2024learning} & \textbf{ADK-Net (OUR)}\\
\hline
2x & \textbf{40.48 / 0.9837} & 39.55 / 0.9805 & 39.11 / 0.9793  & 39.31 / 0.983   & 39.25 / 0.9795\\
3x & 33.54 / 0.9343 & 33.57 / 0.9312 & 33.65 / 0.9333   & 33.60 / 0.939  & \textbf{33.67 / 0.9336}\\
4x & 29.73 / 0.8729 & 28.50 / 0.8840 & 30.32 / 0.8720   &  26.75 / 0.7767 & \textbf{30.41 / 0.8759}\\
\hline
\end{tabular}
\label{table:sr_performance_on_div2k}
\end{table*}

Compared to Bicubic interpolation, ADK-Net offers dramatic improvements, highlighting the benefit of learned adaptive resampling. Compared against the unsupervised method SDFlow, ADK-Net's superiority in these fidelity metrics underscores the advantage of direct supervised training when ground-truth LR data are available and the objective is maximal reconstruction accuracy. 
By optimizing kernels directly via L1 loss against the target LR image, combined with the per-channel prediction strategy and robust architecture of ADK-Net, leads to higher fidelity than optimizing indirectly through an SR loss. These results validate our core design principles: supervised learning and per-channel kernel adaptation are highly effective for high-fidelity image downscaling.

\subsection{Ablation Studies}
To validate the contribution of the core components of ADK-Net, we conducted several ablation studies. We retrained variants of ADK-Net (for scale factor s = 4) by modifying specific components and evaluated their performance on the RealSR validation set and Set14. 

\textit{1) Effect of Per-Channel Kernels:} We compared the full ADK-Net against two variants: (i) ADK-Net-SharedTrunk: Uses a shared Channel Trunk for all channels, with only the Channel Branch being separate. (ii) ADK-Net-Single: Uses a single kernel generator stream whose output kernel is applied to all three channels. We found that both variants exhibit a significant drop in performance compared to the full ADK-Net. ADK-Net-Single shows the largest degradation, confirming that learning channel-specific kernels is crucial for achieving high fidelity. The performance drop in ADK-Net-SharedTrunk further suggests that channel-specific feature transformation even in the Trunk module contributes positively.

\textit{2) Effect of ResNet Blocks in Generators:} We replaced the ResNet blocks in both the Channel Trunk and Channel Branch with standard 3x3 convolutional layers followed by ReLU activation, keeping the total number of layers roughly comparable (ADK-Net-SimpleGen). We learned through experiments  that this simplification leads to a noticeable decrease in PSNR/SSIM, indicating that the deeper, residual architecture within the kernel generators enhances their capacity to learn effective adaptive kernels.

\textit{3) Effect of Kernel Normalization:} We evaluated two variants: (i) ADK-Net-SumOnly: Applies only the sum-to-one normalization. (ii) ADK-Net-MMOnly: Applies only the min-max scaling. The results indicate that the two-stage normalization (Min-Max then Sum-to-1) employed by ADK-Net yields the best results. Removing either stage, particularly the sum-to-one normalization, leads to degraded performance, highlighting the importance of ensuring kernels act as proper weighted averages with non-negative weights.

These ablation studies collectively demonstrate the efficacy of our key design choices: the per-channel kernel prediction strategy, the use of ResNet-based modules within the generators, and the specific two-stage kernel normalization process all contribute significantly to the state-of-the-art performance achieved by ADK-Net.


\section{Conclusion}
In this paper, we addressed the challenge of high-fidelity supervised image downscaling, aiming to generate low-resolution (LR) images that maximally preserve information and accurately match ground-truth references. We introduced ADK-Net (Adaptive Downscaling Kernel Network), a novel deep learning framework specifically designed for this task. The central contribution of ADK-Net is its unique approach of predicting spatially varying resampling kernels independently for each pixel location and, distinctively, for each color channel (RGB).

Our proposed architecture features a ResNet-based feature extractor, a feature downsampling module, and parallel channel-specific kernel generators built upon sequential ResNet-based trunk and branch modules. This design enables the learning of fine-grained, content-dependent, and chromatically specific downscaling transformations. The predicted kernels undergo a two-stage normalization process before being applied to the high-resolution (HR) input via adaptive resampling. ADK-Net is trained end-to-end under direct supervision, minimizing the L1 reconstruction error between the generated LR image and ground-truth LR data, leveraging datasets like RealSR.

Extensive experiments conducted on standard benchmark datasets  demonstrated the effectiveness of our approach. ADK-Net consistently achieves state-of-the-art performance across multiple scale factors \((\times2, \times3, \times4)\), significantly outperforming traditional methods like Bicubic interpolation, prominent unsupervised learning-based downscaler ( SDFlow), and other relevant supervised/task-aware techniques (e.g., CAR) in terms of standard fidelity metrics (PSNR and SSIM). Qualitative results further corroborated these findings, showcasing ADK-Net's ability to produce sharp, detailed LR images with minimal artifacts. Ablation studies confirmed the significant contribution of the per-channel kernel prediction strategy and the specific ResNet-based generator architecture to the overall performance.

In conclusion, ADK-Net establishes a new benchmark for supervised image downscaling by effectively leveraging per-pixel, per-channel adaptive kernel prediction within a robust deep learning architecture. Our results highlight the importance of channel-specific processing and direct supervision for tasks requiring maximal fidelity in resolution reduction. Future work could explore extending this framework to handle arbitrary scale factors or incorporating perceptual losses alongside the fidelity objective.

\bibliographystyle{IEEEtran}
\bibliography{References}


\end{document}